\pdfoutput=1
\documentclass{article}
\usepackage{multicol}
\usepackage[utf8]{inputenc}      
\usepackage{amsmath,amssymb}     
\usepackage{setspace}            
\usepackage{geometry}            
\usepackage{graphicx}  
\usepackage{caption}             
\usepackage{subcaption}
\usepackage{here}
\usepackage[
  backend=biber,
  style=numeric,
  sorting=none,
  doi=false,url=false
]{biblatex}
\renewbibmacro{in:}{}
\AtEveryBibitem{
  \clearfield{volume}
  \clearfield{number}
  \clearfield{pages}
}
\usepackage[colorlinks=true,citecolor=blue]{hyperref}

\title{Extended Factorization Machine Annealing for Rapid Discovery of Transparent Conducting Materials}

\author{
  Daisuke Makino\textsuperscript{1} \quad
  Tatsuya Goto\textsuperscript{1} \quad
  Yoshinori Suga\textsuperscript{2}\\
  \textsuperscript{1}\small Jij Inc, Minato-ku, Tokyo, 108-0023, Japan\\
  \textsuperscript{2}\small Toyota Motor Corporation, Toyota City, Aichi 471-8571, Japan
}

\date{}

\begin{document}
\maketitle

\begin{abstract}
The development of novel transparent conducting materials (TCMs) is essential for enhancing the performance and reducing the cost of next-generation devices such as solar cells and displays. In this research, we focus on the (Al$_x$Ga$_y$In$_z$)$_2$O$_3$ system and extend the FMA framework, which combines a Factorization Machine (FM) and annealing, to search for optimal compositions and crystal structures with high accuracy and low cost. The proposed method introduces (i) the binarization of continuous variables, (ii) the utilization of good solutions using a Hopfield network, (iii) the activation of global search through adaptive random flips, and (iv) fine-tuning via a bit-string local search. Validation using the (Al$_x$Ga$_y$In$_z$)$_2$O$_3$ data from the Kaggle "Nomad2018 Predicting Transparent Conductors" competition demonstrated that our method achieves faster and more accurate searches than Bayesian optimization and genetic algorithms. Furthermore, its application to multi-objective optimization showed its capability in designing materials by simultaneously considering both the band gap and formation energy. These results suggest that applying our method to larger, more complex search problems and diverse material designs that reflect realistic experimental conditions is expected to contribute to the further advancement of materials informatics.
\end{abstract}

\begin{multicols}{2}
\section{Introduction}

In recent years, there has been a strong demand to improve the performance and reduce the costs of optoelectronic devices such as solar cells, flat-panel display LEDs, transistors, sensors, touchscreens, and lasers \cite{Ramprasad2017, MirshojaeianHosseini2021, Li2024, Yang2024}. These devices require transparent conducting materials (TCMs) that simultaneously achieve high electrical conductivity and high transmittance in the visible light region. However, in general, conductivity and optical transparency are properties that conflict with each other, and only a very limited number of compounds can fulfill both simultaneously \cite{WoodsRobinson2024}. In particular, there are few practical examples of p-type TCMs, and even for n-type TCMs, issues remain regarding resource availability and costs, such as dependence on the rare element indium. Against this background, discovering and designing new TCMs is regarded as an important theme directly linked to solving a broad range of problems in energy, environment, society, and economics \cite{Kawazoe1997, Sarma2019, Sutton2019}.

The aluminum (Al), gallium (Ga), and indium (In) ternary sesquioxides that we focus on in this study are promising TCM candidates capable of achieving a large bandgap energy in the visible range (an indicator of high transparency) and high conductivity \cite{Sutton2019, Seacat2021}. These ternary mixed oxides have the composition (Al$_x$Ga$_y$In$_z$)$_2$O$_3$ (where $x+y+z=1$), and by further adjusting dopants and lattice site occupancy, improvements in properties not attainable with conventional binary systems can be expected. However, there are enormous degrees of freedom for actual alloying, making it difficult to comprehensively explore “which composition and structure is optimal.” While first-principles calculations (DFT) can quantitatively determine formation energies and bandgaps, large-scale systematic screening is hindered by extremely high computational costs \cite{Bloechl1994, Kresse1996, Kumagai2023}.

Meanwhile, materials informatics methods that have emerged in recent years open up possibilities to efficiently predict properties using machine learning models built on computational and experimental data, exploring vast compositional spaces in a short time \cite{Ward2016, Fare2022, }. For example, the 2018 Kaggle competition “Nomad2018 Predicting Transparent Conductors” focused on building machine learning models that accurately predict formation energies and bandgaps in (Al$_x$Ga$_y$In$_z$)$_2$O$_3$, and participants competed with various algorithms (graph neural networks, gradient boosting, SOAP descriptors + neural networks), demonstrating high performance \cite{Sutton2020}. Moreover, there has been a rapid expansion of approaches in which such high-accuracy learned models of material properties are treated as “inverse problems” to discover novel material compositions that fulfill targeted properties and functionalities \cite{Balachandran2017, Ren2018}. Specifically, by setting the desired properties in advance, inputting composition and structure into the machine learning model, and iteratively optimizing while evaluating the predicted properties, one can narrow down the vast compositional space, discovering promising materials in far less time than traditional methods. Such research leveraging inverse-problem approaches has also contributed to cost reduction for experiments and simulations and is expected to further accelerate the material development process.

In this study, we conducted searches for undiscovered structures using Factorization Machine Annealing (FMA) on TCM datasets exemplified by Nomad2018, demonstrating its high speed and effectiveness. As in previous works, we use two evaluation metrics—formation energy and bandgap—and our goal is to optimize them simultaneously or assess their trade-offs. Factorization Machines (FM) are known for compactly and efficiently expressing second-order interactions among multiple features \cite{Rendle2010, Rendle2009}, though direct applications in the materials field remain relatively scarce. Moreover, the annealing approach combined as the search algorithm—whether simulated annealing or quantum annealing—has strengths for large-scale combinatorial optimization, but concerns have been raised regarding how to represent continuous variables and the risk of converging to local minima \cite{Kirkpatrick1983, Kadowaki1998, Kitai2020, Xu2025}. In this study, we compare several methods of converting continuous parameters into binary variables and introduce a Hopfield penalty to strengthen exploitation, together with adaptive random flips to enhance exploration. These methods demonstrate more efficient convergence to good solutions than conventional approaches.

Experimental comparisons confirmed that this approach can search far more rapidly than conventional methods using Bayesian optimization or genetic algorithms, thereby discovering promising composition and structure candidates even with limited computational resources. These results support the importance of combining “high-accuracy predictive models” and “fast optimization” in materials design, and are expected to contribute to TCM applications in various areas including energy and environmental issues. In this chapter, we first describe the significance and challenges of TCMs, together with an overview of materials informatics methods. The next chapter provides more detailed previous research, covering machine learning applications to TCM—including (Al$_x$Ga$_y$In$_z$)$_2$O$_3$—and trends in the search algorithms currently in use.

\section{Related Work}

\subsection{Materials Informatics for TCMs}

Because TCMs possess both high transmittance in the visible range and high electrical conductivity, they are seen as a core material for next-generation displays, solar cells, and various sensor devices \cite{Kawazoe1997, Sutton2019, Tokura1989}. Traditionally, n-type oxides containing indium—like ITO (In$_2$O$_3$:Sn)—have been predominant, but in addition to the uneven distribution of resources and rising costs, there remain issues such as the persistent lack of practical p-type materials, posing many challenges in materials development \cite{WoodsRobinson2024, Sarma2019, Kawazoe2000, Brunin2019}. In response to these circumstances, recent materials informatics research has been very active in using high-level computational and experimental data to discover and optimize new TCM compositions and structures \cite{Ramprasad2017, WoodsRobinson2024, Sutton2020}.

For example, numerous neural network models have been reported for predicting bandgap, conductivity, and formation energy; in particular, graph neural networks (CGCNN, MEGNet) that represent crystal structures as graphs have shown high accuracy \cite{Xie2018, Chen2019, Goodall2019}. Ensemble learning, such as gradient boosting decision trees and random forests, has also achieved strong results on various materials science benchmarks. In the top entries of the Nomad2018 Predicting Transparent Conductors competition, crystal-graph NNs, gradient boosting, SOAP descriptor + neural networks, etc., were used, and their average prediction errors approached the accuracy of DFT \cite{Sutton2020}.

\subsection{Kaggle: Nomad2018 Predicting Transparent Conductors}

In 2018, the online platform Kaggle hosted “Nomad2018 Predicting Transparent Conductors,” where around 3,000 data points of (Al$_x$Ga$_y$In$_z$)$_2$O$_3$ were used to build competing models predicting formation energies and bandgaps. Since this dataset contained both the bandgap—an indicator of transparency in the visible region—and the formation energy—an indicator affecting the stability of the material—it held high importance for actual TCM design \cite{Sutton2019, Sutton2020}. In particular, for (Al, Ga, In) mixed crystals, the composition varies continuously within $x + y + z = 1$, and there are many degrees of freedom such as atomic arrangement and lattice constants. Covering such a vast search space with only conventional DFT calculations is unrealistic, making it crucial to have a fast, data-driven predictive model \cite{Bloechl1994}.

The competition’s results showed that the top models, such as crystal graph-based neural networks and gradient boosting, achieved high accuracy using diverse approaches. However, it was also suggested that each model had strengths and weaknesses in certain compositional regions, indicating differences in domain applicability. Subsequent follow-up research has studied ensembles of multiple models, uncertainty evaluation when predicting outside the domain, and so forth. Additionally, this competition stimulated various further studies, including comparisons of machine learning approaches using TCM datasets, knowledge about extrapolating to unknown compositional regions, and new applications \cite{Balachandran2017, Ren2018}.

\subsection{Materials Exploration and Optimization Algorithms}

In addition to merely predicting material properties with machine learning models, the inverse-problem approach of “designing material structures that possess the desired properties” has also attracted attention \cite{ChenLi2021}. Conventionally, genetic algorithms (GA) or particle swarm optimization (PSO) have been used for crystal structure prediction (CSP) and dopant optimization \cite{Chen2017GA, Wang2010}, but in recent years, attention has turned to advanced methods like Bayesian optimization, reinforcement learning, and quantum annealing. Bayesian optimization can efficiently explore parameter space with limited evaluations and has been used in combination with experiments or first-principles calculations \cite{Xue2016, Fukazawa2019, Sakurai2019}. Reinforcement learning has been applied to sequential decision-type materials design (e.g., process control) \cite{Pan2022, Zamaraeva2023, Govindarajan2024, Karpovich2024}, and attempts to apply quantum annealing to discrete and combinatorial optimization for faster solutions are also progressing \cite{Kitai2020,Kim2022, Kim2024, Xu2025}.

Meanwhile, frameworks that use models like Factorization Machines (FM)—which can express sparse features and multiple interactions with few parameters—in conjunction with some form of annealing to find optimal solutions have begun showing results in areas such as metamaterial design. In the research by Kitai et al., a factorization machine was employed as a regression model, and its output was combinatorially optimized using quantum annealing, thereby enabling rapid design of new multilayer structures \cite{Kitai2020}. The novelty of the present study lies in applying this FM–annealing combination to explore the crystal structures of TCM. Moreover, after comparing multiple methods of binarizing continuous parameters, we have introduced a Hopfield penalty to promote exploitation and adaptive random flips to promote exploration.

\subsection{Positioning and Contribution of This Study}

As described above, many machine learning approaches have already been proposed for TCM materials design. However, in the case of systems with extremely wide compositional freedom—such as (Al$_x$Ga$_y$In$_z$)$_2$O$_3$—it is not straightforward to simultaneously achieve high-accuracy property prediction and rapid combinatorial searches. Genetic algorithms and Bayesian optimization each have their advantages, but issues such as compositional continuity or local optimum problems could become barriers. The FMA approach proposed here provides a new angle on this challenge by combining the simplicity of FM with the global search capabilities of annealing. In this paper, starting from Chapter 3, we present the algorithmic background of FMA and details on how it is introduced, followed by a thorough discussion of empirical results using Nomad2018 data.

\section{Methodology}

\subsection{Problem Setting}
In this study, we aim to optimize material properties (e.g., bandgap, formation energy) of (Al$_x$Ga$_y$In$_z$)$_2$O$_3$. Concretely, we determine parameters such as composition ratios $(x,y,z)$, crystal lattice lengths $(a,b,c)$, and angles $(\alpha,\beta,\gamma)$ simultaneously, in order to maximize or minimize the target property. Examples of “constraint conditions” include a composition constraint $(x+y+z) = 1$.

While these property values (e.g., bandgap, formation energy) can theoretically be obtained by first-principles (DFT) calculations, the computational costs for systematic large-scale screening are high. Meanwhile, recent studies have reported that high-accuracy models such as ElemNet or Kaggle-winning solutions can achieve performance comparable to DFT for a variety of material systems \cite{Sutton2019}. Therefore, in this research, we treat such high-accuracy models (ElemNet or top Kaggle solutions) as “black-box functions,” denoted $\hat{y}(\mathbf{x})$. In other words,

\begin{quote}
Black-box function $\hat{y}(\mathbf{x})$ = the predicted property value (bandgap, formation energy, etc.) from ElemNet or Kaggle models
\end{quote}

We then explore for $\mathbf{x}$ that maximizes or minimizes $\hat{y}(\mathbf{x})$. In this chapter, we introduce the method that combines Factorization Machine (FM) and annealing (FMA) as the framework for achieving this, and discuss extended elements such as the method to incorporate constraints (extended Lagrangian) and Hopfield networks.

\subsection{Factorization Machine (FM): Model and Training}

Factorization Machines (FM) were proposed as a model that compactly represents second-order interactions in high-dimensional, sparse feature vectors \cite{Rendle2009}. For an input feature vector $\mathbf{x} \in \mathbb{R}^n$, the FM prediction $\hat{y}(\mathbf{x})$ is defined as:
\[
\hat{y}(\mathbf{x})
= w_0
+ \sum_{i=1}^n w_i \, x_i
+ \sum_{i=1}^n \sum_{j=i+1}^n
\langle \mathbf{v}_i, \mathbf{v}_j \rangle \, x_i \, x_j,
\]
where $w_0$ and $\{w_i\}$ are the bias and linear terms, and each $\mathbf{v}_i \in \mathbb{R}^k$ is a latent vector. Enumerating all second-order interactions would require $O(n^2)$ computations, but by using inner products of latent vectors, FM reduces this to $O(kn)$.

When building a regression model with FM, the following loss function is typically optimized:
\begin{equation*}
\begin{split}
\min_{w_0,\{w_i\},\{\mathbf{v}_i\}} \;&
\sum_{m=1}^M \bigl( y^{(m)} - \hat{y}\bigl(\mathbf{x}^{(m)}\bigr) \bigr)^2 \\
& \quad + \lambda \Bigl(\|\mathbf{w}\|^2 + \sum_i \|\mathbf{v}_i\|^2 \Bigr),
\end{split}
\end{equation*}
where $(\mathbf{x}^{(m)}, y^{(m)})$ is a training data pair, and $y^{(m)}$ is an experimentally measured or computed value such as bandgap or formation energy. The strength of L2 regularization $\lambda$ is adjusted to suppress overfitting. For optimization, one typically employs stochastic gradient methods such as gradient descent or Adam \cite{Kingma2014}, with dynamic learning rate adjustments that can help accelerate and stabilize convergence.

\subsection{QUBO Design for Exploration–Exploitation Balance}

\subsubsection{Bit Representation and Constraints}
Because (Al$_x$Ga$_y$In$_z$)$_2$O$_3$ has continuous parameters such as composition $(x,y,z)$ and lattice constants $(a,b,c,\alpha,\beta,\gamma)$, we must map these into a QUBO (Quadratic Unconstrained Binary Optimization) format to solve by annealing. Specifically, each parameter is first scaled from its original range to the interval $[0,1]$, and then discretized, for instance, by dividing into $2^{n_{\mathrm{bits}}}$ intervals to form a bit string
\[
\mathbf{b} = (b_0,b_1,\dots,b_{n_{\mathrm{bits}}-1}) \in \{0,1\}^{n_{\mathrm{bits}}},
\]
thus discretizing the variable. A standard binary fractional representation
\[
  x(\mathbf{b}) 
  = \frac{\sum_{k=0}^{n_{\mathrm{bits}}-1} b_k \, 2^k}{2^{n_{\mathrm{bits}}}}
\]
allows conversion between $\mathbf{b}$ and continuous $x$, though other encodings such as Gray encoding are also possible. In addition, the constraint $(x + y + z = 1)$ can be incorporated via the extended Lagrangian method \cite{Hestenes1969}:
\[
(x(\mathbf{s}) + y(\mathbf{s}) + z(\mathbf{s}) - 1)^2,
\]
added to the energy function. Here, $\mathbf{s}$ merges the bit strings for all parameters ($\mathbf{b}_x,\mathbf{b}_y,\mathbf{b}_z,\dots$). Decoding $\mathbf{s}$ yields continuous values $\mathbf{x}(\mathbf{s})$.

\subsubsection{Strengthening Exploitation: Hopfield Penalty}
To stabilize good solutions previously found, and leverage them after retraining or changes in constraints, we introduce a Hopfield network \cite{Hopfield1982}. This model sets interactions among $\pm1$-type node states and can store particular bit patterns as energy minima.

By registering the bit string $\mathbf{s}^{(\ell)}$ found in past annealing to the Hopfield network, the penalty $\Psi_{\mathrm{hopfield}}(\mathbf{s})$ used in this study is designed so that $\mathbf{s}$ closer to $\mathbf{s}^{(\ell)}$ has lower energy. Consequently, when re-optimizing, solutions do not deviate too far from past good solutions, producing an exploitation effect. However, relying too heavily on Hopfield can make it difficult to escape local minima, so it must be combined with the random flip approach (to enhance exploration) described next.

\subsubsection{Strengthening Exploration: Adaptive Random Flip}
While Hopfield enhances exploitation, global exploration is equally important. Thus, in this study, we incorporate adaptive random flips during annealing. Each bit $s_i$ is flipped with probability $p$, and when the search stagnates, $p$ is increased; when improvements are observed, $p$ is decreased, thus avoiding being trapped in local minima.

\subsubsection{Energy Function and Optimization Schedule}
To deal with the Hopfield penalty, constraint terms, and the target property $\hat{y}$ to be minimized (or maximized) in a unified manner, we set the following energy function:
\begin{equation*}
\begin{aligned}
\min_{\mathbf{s}} \; E(\mathbf{s})
&= \alpha \, \hat{y}\bigl(\mathbf{x}(\mathbf{s})\bigr)\\
&\quad + \beta \,\Psi_{\mathrm{lagrange}}\bigl(\mathbf{x}(\mathbf{s})\bigr)\\
&\quad + \gamma \,\Psi_{\mathrm{hopfield}}(\mathbf{s}).
\end{aligned}
\end{equation*}

Here, $\mathbf{s}$ is a set of binary variables in $\{0,1\}^M$, and $\hat{y}(\mathbf{x}(\mathbf{s}))$ corresponds to an objective function such as bandgap or formation energy. Depending on the problem, if one wants to treat it as a maximization problem, simply flipping $\alpha\,\hat{y}$ to $-\alpha\,\hat{y}$ will let you handle the problem in a minimization framework. Meanwhile, $\beta\,\Psi_{\mathrm{lagrange}}$ reflects constraints like $(x + y + z - 1)$ through a penalty, and $\gamma\,\Psi_{\mathrm{hopfield}}(\mathbf{s})$ is responsible for stabilizing previously obtained good solutions via the Hopfield network.

\subsection{Overall Framework of the Extended FMA}
By integrating the FM approach, bit representation, constraint incorporation, Hopfield penalty, adaptive random flips, etc., we arrive at an extended FMA framework, expanding on the existing method. Below is the overall flow.

\textbf{(1) Preparation of FM or a Black-Box Function}\\
First, we arrange a setup capable of computing the property values $\hat{y}(\mathbf{x})$. As black-box functions, it is desirable to use low-cost, high-accuracy models such as ElemNet or Kaggle models.

\textbf{(2) Discretization via Binary Encoding}\\
We convert the composition and lattice constants of (Al$_x$Ga$_y$In$_z$)$_2$O$_3$ to bit strings and include constraints like $(x + y + z = 1)$ as penalties via the extended Lagrangian method. In this way, continuous parameters can be treated as QUBO.

\textbf{(3) Incorporation of the Hopfield Network}\\
We register the good solutions found in previous annealing runs into the Hopfield network, using $\Psi_{\mathrm{hopfield}}(\mathbf{s})$ to strengthen exploitation. This step encourages the solutions not to deviate excessively from earlier good solutions, even after retraining.

\textbf{(4) Simulated Annealing}\\
To minimize the energy function $E(\mathbf{s})$, we update the bit string $\mathbf{s}$. We also combine adaptive random flips to promote exploration and avoid stagnation in local minima.

\textbf{(5) Decode, Evaluate, and Local Search}\\
We decode the best solution $\mathbf{s}^*$ into continuous parameters $\mathbf{x}^*$ and evaluate them with the black-box function (ElemNet, Kaggle model, etc.).  
However, because the QUBO learned by FM does not precisely represent the original problem,  
$\mathbf{s}^*$ is not necessarily the true optimal solution.  
Hence, this study also performs local search around $\mathbf{s}^*$, flipping a few bits in its neighborhood to achieve further improvement in the score.  
Even if there is some approximation error from the FM-based QUBO, there might be better solutions near $\mathbf{s}^*$,  
so a bit-based fine-tuning approach is expected to be effective \cite{Endo2024}.

By iterating this process, one can efficiently solve a continuous-parameter exploration problem with multiple constraints using simulated annealing.  
Additionally, the procedure of performing bit-level local searches on the QUBO optimum from FM can compensate for small discrepancies with the original problem,  
helping discover even higher-quality solutions.  
In the next chapter, we present numerical experimental results using this method and examine its effectiveness from various perspectives.

\section{Experimental Results and Discussion}

In this chapter, we present the results of numerical experiments using the extended FMA (Factorization Machine + Annealing + Local Search) described in Chapter 3.  
First, we address a single-objective optimization problem of maximizing the bandgap, using the (Al$_x$Ga$_y$In$_z$)$_2$O$_3$ dataset provided by Kaggle’s “Nomad2018 Predicting Transparent Conductors” \cite{Sutton2019},  
and compare our method’s performance against Bayesian optimization (TPE).  
Next, we investigate the performance of extended FMA from multiple viewpoints. Specifically,  
we examine the impact of the Hopfield penalty coefficient on the search performance,  
and verify how the performance differs when changing the encoding method (binary or Gray code) or the number of bits (4 bits, 8 bits, 16 bits, etc.) for converting continuous variables into binary.  
Through these investigations, we clarify in detail whether the new ideas introduced in extended FMA (Hopfield penalty, adaptive random flips, and local search) function effectively.  
Finally, we introduce a multi-objective optimization scenario treating both bandgap and formation energy in (Al$_x$Ga$_y$In$_z$)$_2$O$_3$,  
testing the possibility of more flexible optimization.  
We also take up the MaxCut problem as an application example beyond (Al$_x$Ga$_y$In$_z$)$_2$O$_3$,  
showing that the mechanisms adopted by extended FMA are generally valid in other domains as well.

\subsection{Experimental Setup}

\noindent
\textbf{(1) Comparisons of Extended FMA with Other Methods}\\
As the first stage of experiments, we use the (Al$_x$Ga$_y$In$_z$)$_2$O$_3$ dataset from the Kaggle competition “Nomad2018 Predicting Transparent Conductors” \cite{Sutton2019}.  
We employ an ElemNet model trained on this dataset as our black-box function,  
estimating the bandgap $\hat{E}_{\mathrm{gap}}(\mathbf{x})$ from inputs such as composition $(x,y,z)$ and lattice constants $(a,b,c,\alpha,\beta,\gamma)$.  
Our task is single-objective optimization to maximize this quantity.  
We first examine how extended FMA performs in comparison with Bayesian optimization (TPE).

\vspace{1em}
\noindent
\textbf{(2) Verifying Extended FMA’s Performance}\\
Next, we verify extended FMA’s search performance from multiple angles.  
Specifically, we vary the Hopfield penalty coefficient to evaluate the stability of optimization and the ability to escape local minima.  
We also systematically examine the performance differences when changing the encoding format (binary vs. Gray code) or the bit length (4 bits, 8 bits, 16 bits, etc.) for converting continuous variables into binary.  
Additionally, we confirm the extent to which introducing local search improves search accuracy, clarifying the effectiveness of the novel ideas introduced in extended FMA.

\vspace{1em}
\noindent
\textbf{(3) Multi-Objective Optimization}\\
In this study, we also attempt a multi-objective optimization scenario that deals simultaneously with bandgap $(E_g)$ and formation energy $(E)$.  
Concretely, we set up an exploration where $E_g$ is brought close to a certain value while minimizing $E$,  
demonstrating the possibility that extended FMA can adjust multiple properties at once instead of just a single objective.  
Moreover, we investigate whether it can search for property values not actually present in the data (e.g., a bandgap of 4.0 eV),  
thus examining the flexibility of the inverse-problem approach with extended FMA.

\vspace{1em}
\noindent
\textbf{(4) Application to the MaxCut Problem}\\
Finally, as a general optimization problem different from the (Al$_x$Ga$_y$In$_z$)$_2$O$_3$ system, we take up the MaxCut problem and verify whether Hopfield penalties and adaptive random flips also contribute to global exploration in non-materials domains.  
In doing so, we suggest that the concepts and implementation of extended FMA can be applied to a wide range of optimization tasks,  
indicating future directions for development.

\subsection{Comparison of Extended FMA with Other Methods}

In this section, we compare three methods—extended FMA, TPE, and NSGA-II—by running each for 60 seconds of optimization and evaluating the final best score obtained within that period. TPE and NSGA-II are implemented using the Bayesian optimization framework Optuna\cite{Akiba2019}.

\begin{figure}[H]
 \centering
 \includegraphics[width=\linewidth]{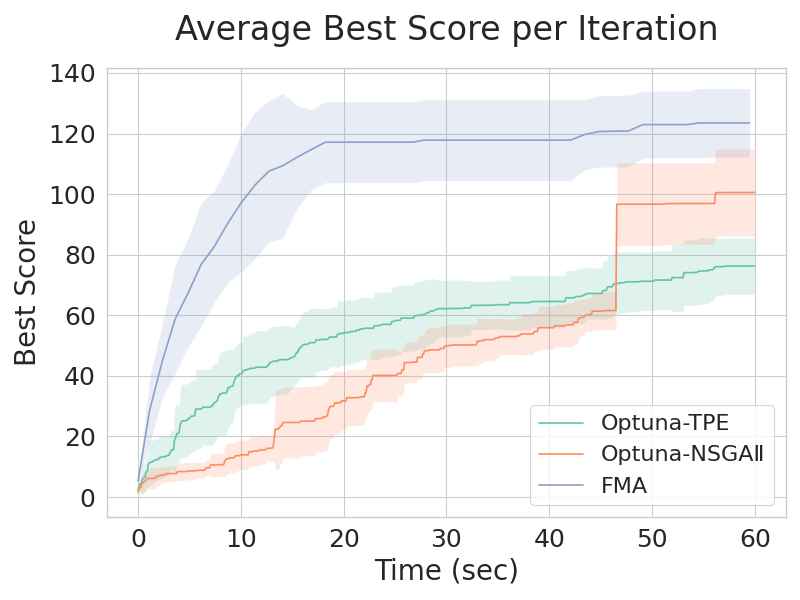}
 \caption{Comparison among FMA, NSGA-II, and TPE}
 \label{fig:tcm_fma_optuna_comparison}
\end{figure}

As a result, extended FMA achieved the highest score at 123.5, followed by NSGA-II at 100.5, and TPE at 76.2 (Fig.~\ref{fig:tcm_fma_optuna_comparison}). Notably, extended FMA rapidly found major improvements early on due to its update rules, and at the 60-second mark, its average score far exceeded those of the other two approaches. Specifically, it reached NSGA-II’s 60-second score in about 10 seconds and matched TPE’s 60-second score in about 5 seconds.

From these results, for the bandgap maximization problem in the (Al$_x$Ga$_y$In$_z$)$_2$O$_3$ system, extended FMA shows superior search performance compared to the other two methods.  
In the following sections, we further investigate how this proposed method (extended FMA) achieves its results by examining (1) Hopfield penalty coefficients, (2) the number of binary variables, and (3) encoding methods for continuous values.

\subsection{Performance Comparison by Hopfield Penalty Coefficients}

In this section, we present the results of verifying how optimization performance changes when varying the Hopfield penalty coefficient $\lambda$. We tested $\lambda=0.0$, $\lambda=0.5$, and $\lambda=5.0$, each under the same problem settings and search conditions. For all, we ran the process up to iteration 50 and compared the final scores.

\begin{figure}[H]
  \centering
  \includegraphics[width=\linewidth]{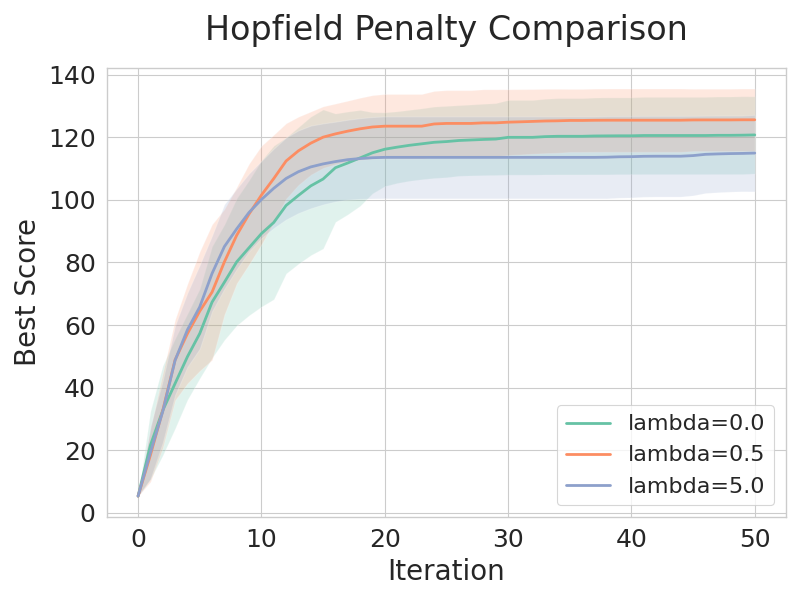}
  \caption{Average score transitions at each iteration for Hopfield penalty coefficients $\lambda=0.0, 0.5, 5.0$.
    The horizontal axis is the iteration, the vertical axis is the score.
    $\lambda=0.5$ reached the highest final value.}
  \label{fig:hopfield_lambda_sidebyside}
\end{figure}

Figure~\ref{fig:hopfield_lambda_sidebyside} and Table~\ref{tab:hopfield_lambda_comparison} show the average score transitions by iteration and the numerical values at iteration 50, respectively. For $\lambda=0.0$ (no penalty), the score reached around 100 by iteration 15, but its growth slowed thereafter, converging at around 120. By contrast, with $\lambda=0.5$, it reached about 100 by iteration 10 and continued to rise, ultimately getting to around 125. Meanwhile, with $\lambda=5.0$, although the early growth was not bad, improvement in the middle and later stages was minimal, ending around 114–115.

These results suggest that when $\lambda=0.0$ (no penalty), there is no mechanism to emphasize good solutions, so the search relies on local exploration only. This can reach a moderate score quickly but makes it difficult to achieve breakthroughs in the later stages. Conversely, $\lambda=5.0$ might be too large, strongly anchoring the search to once-found solutions, weakening global exploration and making it difficult to attain higher scores. A moderate value of $\lambda=0.5$ effectively exploits good solutions without overly constraining exploration, thus ultimately reaching the highest score.

Overall, a large Hopfield penalty coefficient may guide the search strongly toward certain good solutions but narrow the scope of exploration, hindering further improvements. On the other hand, not using it at all ($\lambda=0.0$) may allow local-search-based improvements but misses opportunities for major breakthroughs later. In our experiment, $\lambda=0.5$ gave the highest score, suggesting that an appropriate penalty value strikes a balance between “utilizing good solutions” and “global improvement.”

\begin{table}[H]
  \centering
  \caption{Comparison of average scores at iteration 50}
  \label{tab:hopfield_lambda_comparison}
  \begin{tabular}{|c|c|c|}
    \hline
    label & iteration & average score \\
    \hline
    $\lambda=0.0$ & 50 & 120.78 \\
    \hline
    $\lambda=0.5$ & 50 & 125.61 \\
    \hline
    $\lambda=5.0$ & 50 & 114.97 \\
    \hline
  \end{tabular}
\end{table}

\subsection{Relationship between Bit Discretization and Performance}

\begin{figure}[H]
 \centering
 \includegraphics[width=\linewidth]{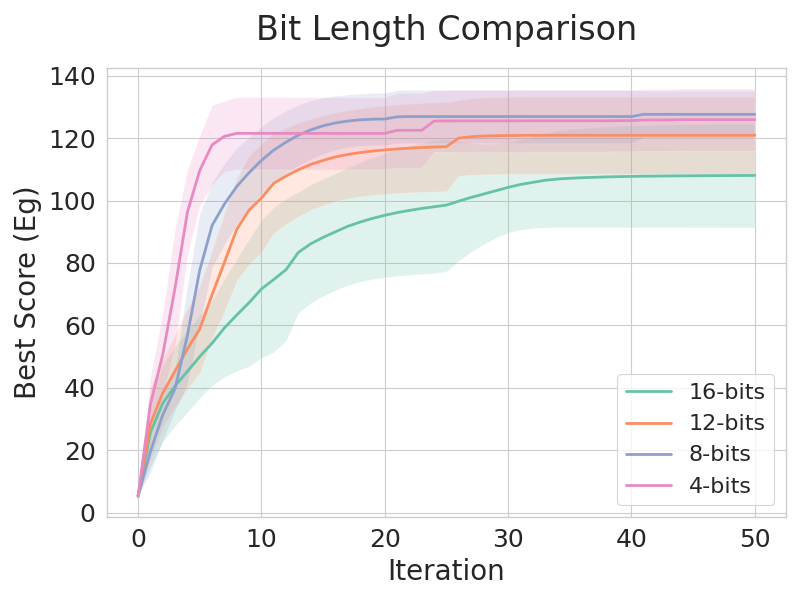}
 \caption{Example of performance transitions when varying the bit length (4, 8, 12, 16)}
 \label{fig:bit_comparison}
\end{figure}

Here, we compare optimization performance when converting continuous parameters into 4-bit, 8-bit, 12-bit, or 16-bit representations. Figure~\ref{fig:bit_comparison} shows the final score transitions for each bit setting. The best final score came from the 8-bit representation, followed by 4-bit, 12-bit, and 16-bit in descending order.

Initially, 4 bits showed a rapid rise in score, but it was overtaken by 8 bits in the mid-to-late phases. This could be because with 4 bits, discretization is coarser, allowing a swift, broader initial search. However, 8 bits, though slightly slower initially, can represent parameters more precisely, thus surpassing 4 bits in final performance after enough iterations.

Meanwhile, with 12 bits and 16 bits, although representational capacity increases, the search space expands, slowing the progress under the current extended FMA strategy. Consequently, final scores do not reach those of 8 bits. It suggests that if bit counts are increased further, one must ensure a sufficient number of iterations and an effective exploration strategy to fully tap that potential.

Hence, an 8-bit scheme appears to balance representational resolution and search efficiency well, yielding the highest final scores in our experiments. Further investigations may reveal ways to optimize algorithm parameters and sampling strategies to draw out even better performance from more bits.

\subsection{Comparison of Different Encoding Methods}

\begin{figure}[H]
 \centering
 \includegraphics[width=\linewidth]{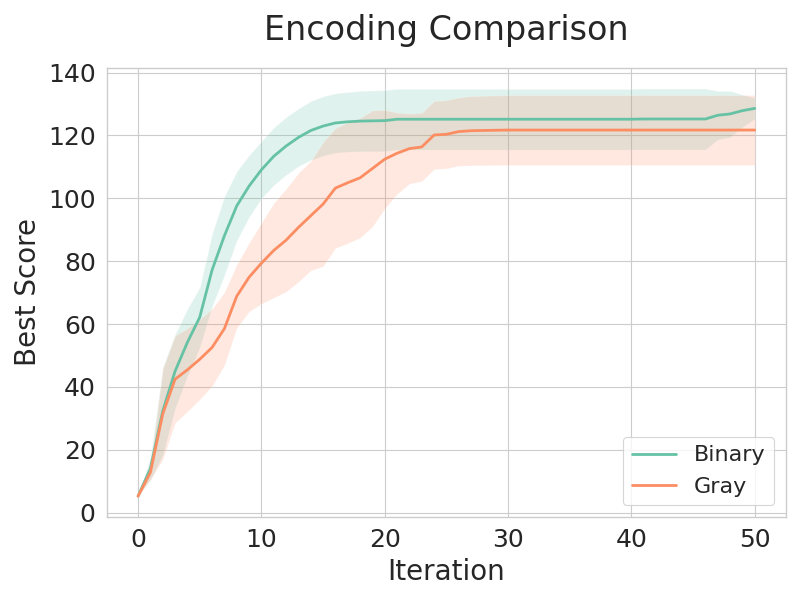}
 \caption{Differences in optimization performance by encoding method (Binary vs. Gray)}
 \label{fig:encoding_comparison}
\end{figure}

In this section, we examine how the choice of encoding method for converting continuous values to binary variables affects optimization performance. Specifically, we compare a simple binary encoding against Gray code. We run each for 50 iterations and evaluate the final scores. Figure~\ref{fig:encoding_comparison} depicts the score transitions for both encoding methods.

From Figure~\ref{fig:encoding_comparison}, the simple binary encoding achieves a slightly higher final score than the Gray code. In our experiment, the final score under binary encoding slightly exceeded that of Gray code. The crucial point here is that the encoding method for continuous parameters does influence extended FMA’s performance, implying that more sophisticated encoding approaches could further enhance extended FMA’s performance.

\subsection{Application to Multi-Objective Optimization}

This subsection presents a simple example of applying extended FMA to multi-objective optimization.  
Specifically, we aim to adjust the bandgap $(E_{\mathrm{g}})$ around 4.0\,eV, while keeping the formation energy $(E)$ as low as possible.  
We chose $E=0.0\,\mathrm{eV}$ and $E_{\mathrm{g}}=4.0\,\mathrm{eV}$ to investigate whether the approach can search for property values not originally present in the dataset.

\begin{figure}[H]
 \centering
 \includegraphics[width=\linewidth]{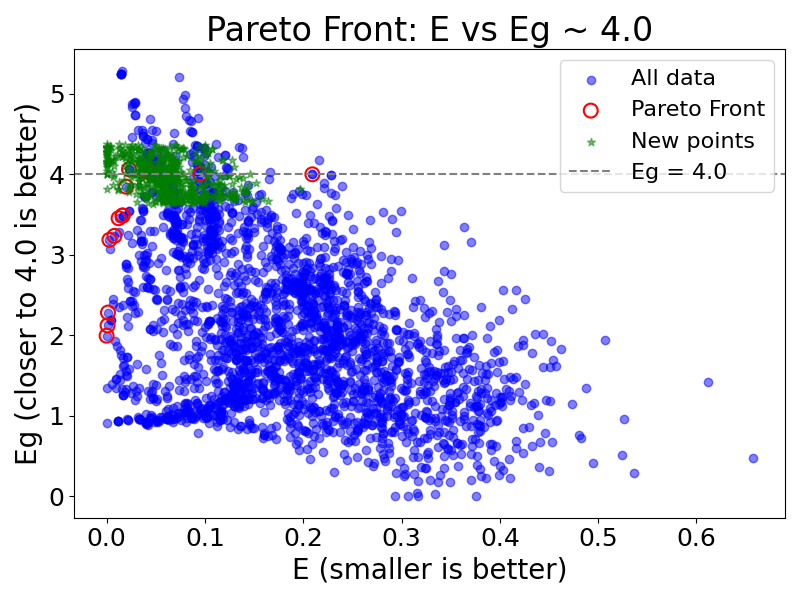}
 \caption{Example of multi-objective optimization results. Red circles represent the Pareto front, green stars denote newly discovered candidates,
   and the gray dashed line indicates \( E_{\mathrm{g}} = 4.0\,\mathrm{eV} \).}
 \label{fig:multi_obj_results}
\end{figure}

Figure~\ref{fig:multi_obj_results} shows the results after optimization.  
Red circles represent the Pareto front derived from existing data,  
and green stars mark candidate points discovered by extended FMA.

Looking at the figure, we see that multiple points have been found where $E$ (formation energy) is kept low while the bandgap $E_{\mathrm{g}}$ is guided near 4.0\,eV.  
This demonstrates that the continuous-parameter exploration by extended FMA can handle not only single-goal maximization/minimization but also multi-objective constraints guiding the solution toward specified regions.  
Future prospects include exploring more realistic setups, multiple properties such as carrier concentration, and controlling specific materials aspects (e.g., reducing the rare element In). This suggests expanding multi-objective optimization further.

\subsection{Validating Random Flips and Local Search in the MaxCut Problem}

The objective of this experiment is to verify whether the Hopfield penalty, random flips, and local search introduced in extended FMA function effectively in problems other than material searches for (Al$_x$Ga$_y$In$_z$)$_2$O$_3$.  
As an example, we applied extended FMA to the MaxCut problem and evaluated its performance in comparative experiments.

In this experiment, we prepared four conditions by combining the presence or absence of random flips (called “flips”) and local search:

\begin{enumerate}
 \item Flip-LocalSearch
 \item Flip-NoLocalSearch
 \item NoFlip-LocalSearch
 \item NoFlip-NoLocalSearch
\end{enumerate}

All four conditions were applied to the same MaxCut problem, iterating the solution update while tracking the cut value (score).

\begin{figure}[H]
    \centering
    \includegraphics[width=\linewidth]{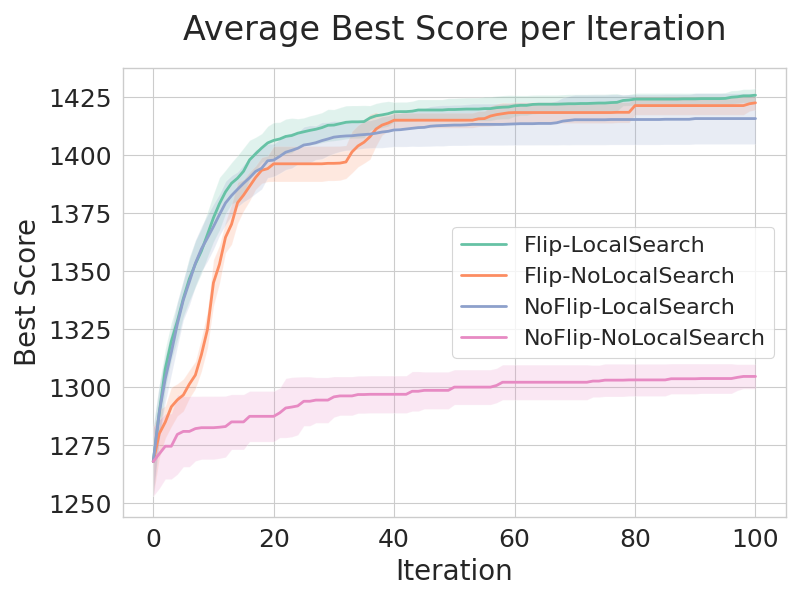}
    \caption{Score transitions for four conditions in the MaxCut problem:
    (Flip-LocalSearch, Flip-NoLocalSearch,
    NoFlip-LocalSearch, NoFlip-NoLocalSearch).
    The horizontal axis is the iteration, and the vertical axis is the average score.
    Flip-LocalSearch achieves the highest final score.}
    \label{fig:maxcut_flip_localsearch}
\end{figure}

Figure~\ref{fig:maxcut_flip_localsearch} shows how each of the four conditions evolves in average score as iterations proceed. By the end (iteration=100), the best score was achieved under Flip-LocalSearch, showing both rapid initial convergence and strong final refinement, outpacing the others. The second-best was Flip-NoLocalSearch, which started slower than NoFlip-LocalSearch but caught up in mid-to-late iterations to finish second.

Interestingly, NoFlip-LocalSearch obtained a high score early on but was overtaken by a flip-based approach around iteration 40, ultimately taking third place. This might be because local search alone was efficient near the initial solution, yet lacking global exploration by flips hindered further improvement in the later stage. Finally, NoFlip-NoLocalSearch produced a very slow increase in scores, finishing with the lowest result.

Overall, global exploration by flips and local exploration by local search complement one another, and combining both (Flip-LocalSearch) yields the highest final score. Meanwhile, the NoFlip-LocalSearch strategy quickly converges to a good solution but its score gain stagnates mid-to-late, eventually being surpassed by Flip-NoLocalSearch. This behavior suggests that maintaining a balance between global (flip) and local (local search) strategies is effective.

\section{Conclusion and Future Outlook}

\subsection{Conclusion}

In this study, we focused on TCMs centered around (Al$_x$Ga$_y$In$_z$)$_2$O$_3$,  
and extended an existing Factorization Machine (FM) + annealing method (FMA),  
introducing new elements such as discretizing continuous parameters, Hopfield penalties,  
adaptive random flips, and bit-based local search. This extended FMA uses a QUBO learned by the FM model to explore parameter space rapidly,  
while leveraging the Hopfield network to utilize past good solutions,  
combining global exploration via random flips and local fine-tuning via local search to achieve efficient and accurate optimization.

From the experimental results presented in Chapter 4, we can summarize the following:

\begin{itemize}
 \item In single-objective optimization (maximizing bandgap), we found that extended FMA can reach high scores earlier than conventional methods (TPE, NSGA-II), discovering good solutions in a short time.
 \item Adjusting factors like the Hopfield penalty coefficient, the number of bits for discretization, and the encoding method significantly affects search performance, suggesting that the novel ideas introduced in extended FMA play an important role in enhancing stability and global exploration.
 \item The method can be easily applied to multi-objective optimization (e.g., bandgap and formation energy). We confirmed that it effectively explored even property values not present in the dataset (e.g., $E_{\mathrm{g}}=4.0\,\mathrm{eV}$).
 \item Extended FMA is also applicable to problems outside materials exploration, such as the MaxCut problem, and we confirmed that the Hopfield penalty and random flips can be effective for global optimization in a general sense.
\end{itemize}

Thus, extended FMA proves to be a powerful method that enables short-time, high-accuracy optimization, capable of having a significant impact on exploring materials systems with vast design freedom, such as (Al$_x$Ga$_y$In$_z$)$_2$O$_3$.

\subsection{Future Outlook}

Based on the experiments verifying extended FMA, the following potential directions for further development emerge:

\begin{itemize}
 \item \textbf{Simultaneously optimizing a wider variety of material properties:}\\
       In this study, we attempted multi-objective optimization for bandgap and formation energy, but in reality, multiple factors—carrier concentration, mobility, film thickness, process conditions, and others—affect material performance. Extending the scenario to optimize these simultaneously could yield solutions with greater practical utility.
 \item \textbf{Improving search efficiency:}\\
       The Hopfield penalty, random flips, and local search can potentially achieve even greater effectiveness with parameter tuning. Various strategies such as optimizing their combinations or adaptive scheduling could reduce computation time for large-scale problems.
 \item \textbf{Integrating materials data:}\\
       Recently, there is strong emphasis on integrating with materials databases and experimental data. By collaborating with high-quality data, one could experimentally validate the approach’s results, incorporating feedback to form a loop that further accelerates development.
\end{itemize}

With these directions in mind, extended FMA is not merely a materials exploration technique but can serve as a general multi-objective optimization framework, contributing to a wide range of fields. Moving forward, by applying it to even larger, more complex search problems and exploring applications reflecting realistic experimental conditions and diverse material designs, it is expected to further advance the progress of materials informatics.


\printbibliography 
\end{multicols}
\end{document}